\documentclass[article,draft,noplatex]{reoroji}  
\usepackage{amsmath,overpic}
\usepackage{bm}  
\begin{document}

\journalyear{2099}
\journalvol{999}
\journalnum{99}
\pagestart{999}
\pageend{999}
\receivedyear{2099}
\receivedmonth{12}
\receivedday{31}

\titleE{Relation between extensional viscosity and polymer conformation in dilute polymer solutions}
\authorE{Yusuke {\sc Koide}${}^{*,\dagger}$, Takato {\sc Ishida}${}^*$, Takashi {\sc Uneyama}${}^*$, Yuichi {\sc Masubuchi}${}^*$}
\authorEshort{KOIDE$\cdot$ISHIDA$\cdot$UNEYAMA$\cdot$MASUBUCHI}

\affiliation{%
${}^*$Department of Materials Physics, Graduate School of Engineering, Nagoya University,\\ Furo-cho, Chikusa, Nagoya, Aichi 464-8603, Japan
}
\correspondingEmail{E-mail: koide.yusuke.k1@f.mail.nagoya-u.ac.jp}
\correspondingTel{Tel: +81-52-789-4202}
\abstract{%

We investigate extensional viscosity and polymer conformation in dilute polymer solutions under uniaxial extensional flow using dissipative particle dynamics simulations.
At high extension rates, polymers are significantly stretched by extensional flows, and the extensional viscosity growth function exhibits strain hardening.
To reveal their quantitative relation, we adopt an analysis method based on the Rouse-type model.
We demonstrate that the extensional viscosity growth function is determined by the instantaneous gyration radii in the parallel and perpendicular directions to the extensional direction and their time derivatives.
Our approach also provides a unified description of the steady-state extensional viscosity of dilute polymer solutions for various chain lengths and concentrations in terms of the polymer gyration radius.
}

\maketitle

\section{INTRODUCTION}

Under fast flows, polymers exhibit significant stretching and alignment, thereby altering the fluid properties even in the dilute regime.
For example, a small amount of polymer can suppress turbulence and reduce friction drag at high Reynolds numbers~\cite{White2008-gb}, whereas at low Reynolds numbers, polymers can instead promote elastic instabilities, leading to chaotic flows~\cite{Steinberg2021-lz}.
From a rheological perspective, the shear viscosity of dilute polymer solutions is almost independent of the shear rate~(with only slight shear thinning), whereas their extensional viscosity increases at high extension rates.
Accordingly, several studies have attempted to relate the drag reduction in turbulent flows to the increase in the extensional viscosity~\cite{lumley1969drag,Dimitropoulos1998-tl,Escudier1999-la,Serafini2022-ot}.

To evaluate the characteristic responses of dilute solutions under extensional flow, a variety of rheological measurement techniques have been developed.
Recently, capillary-driven thinning of a fluid filament has been widely used to realize extensional flows, such as the capillary breakup extensional rheometer~(CaBER)~\cite{McKinley2000-ky,Anna2001-bn} and the dripping-onto-substrate~(DoS)~\cite{Dinic2015-ec}.
Owing to advances in measurement techniques, intriguing properties of dilute polymer solutions under extensional flow have been extensively investigated, including strain hardening and concentration dependence of the apparent relaxation time~\cite{Anna2001-bn,Clasen2006-rc,Tirtaatmadja2006-pa,Dinic2015-ec,Matsuda2024-av,Calabrese2024-dw,Calabrese2025-vc}.
However, several studies have reported difficulties in interpreting results from the capillary-driven thinning due to the dependence on the plate diameter in CaBER~\cite{Gaillard2024-ds,Wang2025-df} and the effect of a moving contact line on the pinch-off dynamics in DoS~\cite{Wu2020-vd}.
Another inherent issue with this type of measurements is that the applied extension rate is self-selected by force balance~\cite{Calabrese2025-vc}, making it difficult to systematically investigate $\dot{\epsilon}$ dependence of extensional rheology.
Cross-slot devices are also used to generate extensional flows.
Recently, Haward et al.~\cite{Haward2023a} developed the optimized uniaxial and biaxial extensional rheometer~(OUBER), a three-dimensional microfluidic device that generates nearly ideal uniaxial and biaxial extensional flows.
In subsequent work, Haward et al.~\cite{Haward2023b} demonstrated that $\dot{\epsilon}$ dependence of the steady-state extensional viscosity of dilute polymer solutions was well described by the finitely extensible nonlinear elastic model with Peterlin closure~(FENE-P), although care must be taken for the flow modulation by polymers at high extension rates.

Complementary use of molecular simulations is indispensable for a molecular-level understanding of the extensional rheology of dilute polymer solutions.
There have been numerous works conducting Brownian dynamics simulations of single-chain models under uniaxial extensional flow~\cite{van-den-Brule1993-mg,Doyle1997-tq,Herrchen1997-fb,Li2000-da,Somasi2002-ot,Hsieh2003-oc,Larson2005-ef}.
Specifically, a drastic polymer stretching and a sharp increase in the extensional viscosity have been investigated extensively to clarify the coil-stretch transition of polymers~\cite{De_Gennes1974-md}. 
The generalized Kraynik--Reinelt~(GKR) boundary condition, recently developed by Dobson~\cite{Dobson2014-kr} and Hunt~\cite{Hunt2016-vl}, enables multi-chain simulations under uniaxial extensional flow at arbitrarily large strains.
Although this approach has been applied to linear and ring polymer melts~\cite{O-Connor2018-mh,OConnor2019-po,O-Connor2020-ez,Murashima2021-wh,Murashima2022-kd}, few studies have addressed dilute polymer solutions under uniaxial extensional flow.
Jana and Dalal~\cite{Jana2025-sh} used the dissipative particle dynamics~(DPD) method to simulate dilute polymer solutions under uniaxial extensional flow, and examined the effect of DPD parameters on the stretching behavior of polymers.
However, they focused only on the end-to-end distance of polymers.

This study aims to reveal the quantitative relation between extensional viscosity and polymer conformation in dilute polymer solutions.
For this purpose, we conduct DPD simulations of dilute polymer solutions under uniaxial extensional flow.
We systematically evaluate the extensional viscosity and the polymer gyration radius, including transient and steady-state behavior.
A key feature of our approach is that we rely on the Rouse-type model, for which a rigorous relation can be derived, to quantitatively connect the rheological properties with polymer conformation.
We demonstrate that the polymer contribution to the extensional viscosity of dilute polymer solutions is described by the gyration radii of polymers in the parallel and perpendicular directions to the extensional direction.

\section{METHOD}
\begin{figure}
  \centering
  \begin{overpic}[width=0.75\linewidth]{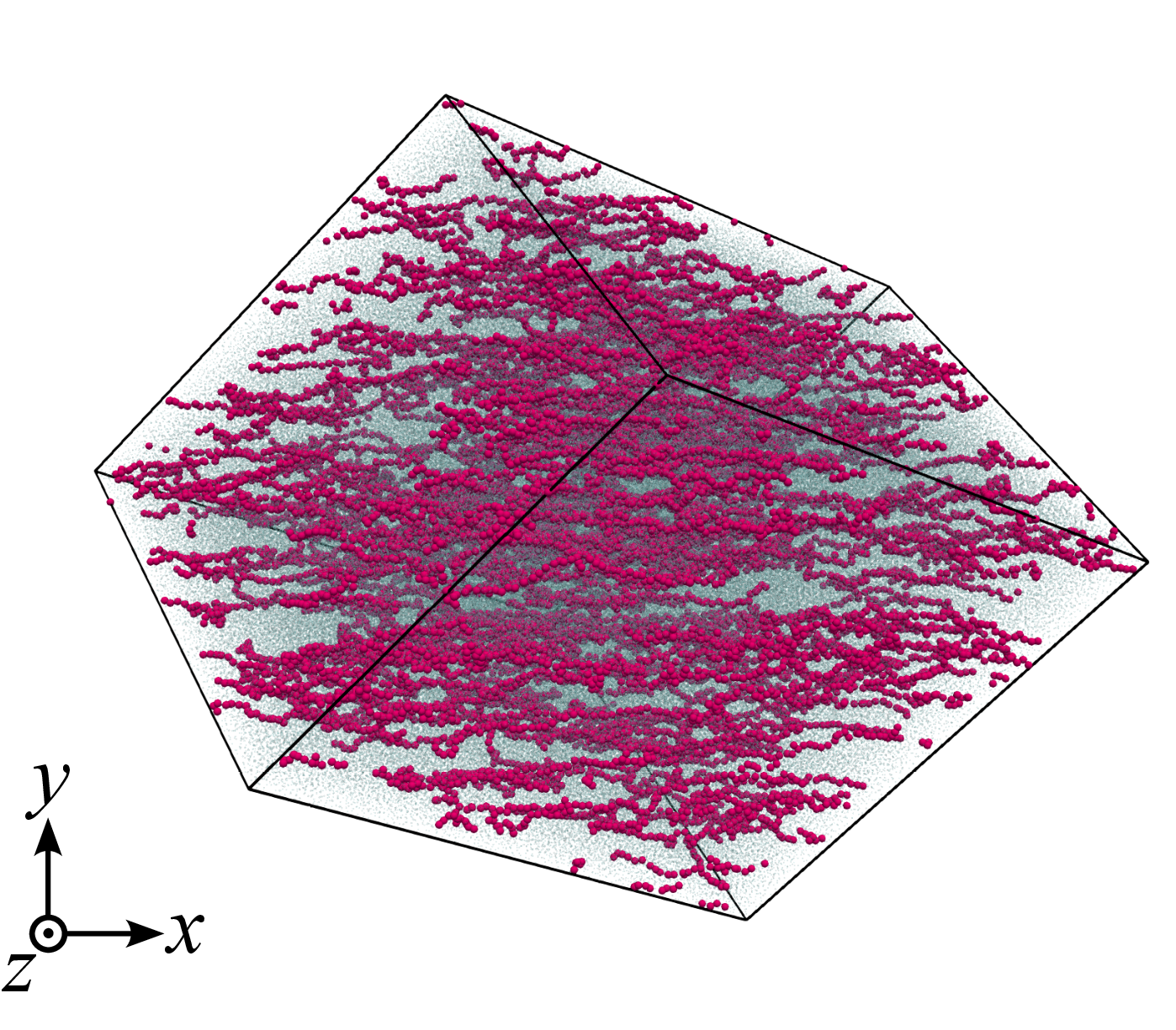} 
  \end{overpic}
  \caption{Snapshot of the polymer solution for $N_\mathrm{p}=50$ and $\phi=0.025$ under uniaxial extensional flow with $\dot{\epsilon}=0.02$. Polymer particles are shown in red. For clarity, solvent particles are represented by blue dots.}
  \label{fig:snapshot}
\end{figure}

We conduct DPD simulations of polymer solutions under uniaxial extensional flow, as shown in Fig.~\ref{fig:snapshot}.
Here, polymers are modeled as fully flexible linear chains, following previous studies~\cite{Jiang2007-rf,Koide2023-ao}.
Each polymer consists of $N_\mathrm{p}$ particles, and adjacent particles are connected by finitely extensible nonlinear elastic~(FENE) springs.
The bonded force $\bm{F}^\mathrm{B}_{ij}$ is given by
\begin{equation}
  \bm{F}^\mathrm{B}_{ij} =-\frac{k_F(|\bm{r}_{ij}|-r_\mathrm{eq})}{1-\{(|\bm{r}_{ij}|-r_\mathrm{eq})/(r_\mathrm{max}-r_\mathrm{eq})\}^2}\bm{e}_{ij},
\end{equation}
where $k_F$ is the spring constant, $r_\mathrm{eq}$ is the equilibrium bond distance, $r_\mathrm{max}$ is the maximum bond distance, $\bm{r}_{ij}=\bm{r}_i-\bm{r}_j$, and $\bm{e}_{ij}=\bm{r}_{ij}/|\bm{r}_{ij}|$ with $\bm{r}_i$ being the position of the $i$-th particle.
In the DPD method, solvent particles are treated explicitly.
These DPD particles interact via repulsive, dissipative, and random forces.
Further details of the DPD method can be found in previous publications~\cite{Koide2022-bp,Koide2023-ao}.
In the following, all quantities are nondimensionalized by $k_BT$, $m$, and $r_c$, where $k_B$ is the Boltzmann constant, $T$ is the reference temperature, $m$ is the mass of a DPD particle, and $r_c$ is the cutoff distance.

In the present study, the parameters of DPD simulations are set as follows:
the total number of particles is $N=648\,000$; the number density of particles is $\rho=3$; the random force coefficient is $\sigma=3$; the spring constant is $k_F=40$; the equilibrium bond distance is $r_\mathrm{eq}=0.7$; the maximum bond distance is $r_\mathrm{max}=2$; the repulsive force coefficients between different types of particles are $a_\mathrm{pp}=25$, $a_\mathrm{ps}=25$, and $a_\mathrm{ss}=25$, where p and s denote polymer and solvent particles, respectively.
Due to the soft-core potential employed in the DPD method, our simulations do not capture entanglement effects~\cite{Pan2002-oq}.
This study investigates the extensional rheology of dilute polymer solutions for $N_\mathrm{p}=25$, $50$, and $80$.
We vary the volume fraction $\phi$ of polymer particles within the dilute regime $\phi<\phi^*$, where the overlap volume fraction $\phi^*$ is defined as $\phi^* = {N_\mathrm{p}}/\{(4/3)\pi \rho R_{g}^3\}$ with $R_{g}$ the polymer gyration radius evaluated at low $\phi(\leq 0.025)$.
In our DPD simulations, the values of $\phi^*$ are $0.24$, $0.14$, and $0.10$ for $N_\mathrm{p}=25$, $50$, and $80$, respectively.

The SLLOD equations~\cite{evans_morriss_2008} and the GKR boundary conditions~\cite{Dobson2014-kr,Hunt2016-vl} are used to generate homogeneous uniaxial extensional flow at arbitrarily large strains.
This approach has been theoretically justified by Daivis and Todd~\cite{Daivis2006-pq,Todd2007-ps}.
The SLLOD equations read 
\begin{align}
  &\frac{d{\bm{r}_i}}{dt} = {\bm{p}_i} + (\nabla \bm{u})^\mathsf{T} \cdot\bm{r}_i \\
  &\frac{d{\bm{p}}_i}{dt} = \bm{F}_i - (\nabla \bm{u})^\mathsf{T} \cdot\bm{p}_i ,
\label{eq:SLLOD_equation}
\end{align}
where $\bm{p}_i$ is the so-called peculiar momentum of the $i$-th particle, $\bm{F}_i$ is the force acting on that particle, and $(\nabla\bm{u})_{ij}=\partial u_j/\partial x_i$ is the velocity gradient tensor with $\bm{u}$ representing the velocity of the applied flow field.
Calculation of the dissipative force, which depends on the relative velocity of interacting particles, is performed using the laboratory velocity.
For uniaxial extensional flow, $\nabla\bm{u}$ is expressed as 
\begin{equation}
  \nabla\bm{u} = \begin{pmatrix}
      \dot{\epsilon}&0&0\\
      0&-\dot{\epsilon}/2&0\\
      0&0&-\dot{\epsilon}/2
  \end{pmatrix},
  \label{eq:uniaxial}
\end{equation}
where $\dot{\epsilon}$ is the extension rate.
Under this extensional flow, the $i$-th lattice basis vector $\bm{e}_i$ of the simulation box evolves as 
\begin{equation}
  \frac{d}{dt}\bm{e}_i = (\nabla \bm{u})^\mathsf{T} \cdot \bm{e}_i\label{eq:box_deform}.
\end{equation}
During simulations of uniaxial extensional flow, we systematically remap the deformed simulation box using the GKR method~\cite{Dobson2014-kr,Hunt2016-vl} and Semaev's algorithm~\cite{Semaev2001-sn}.
To prevent instabilities caused by the finite numerical precision~\cite{Todd2000-rh}, the total peculiar momentum is periodically reset to zero, as in previous studies~\cite{Hunt2016-vl,Nicholson2016-aj}.
We adopt the dynamic size cell list algorithm to reduce the computational cost of short-range interactions under uniaxial extensional flow~\cite{Dobson2016-zk}.

\begin{figure}
  \centering
  \begin{overpic}[width=0.75\linewidth]{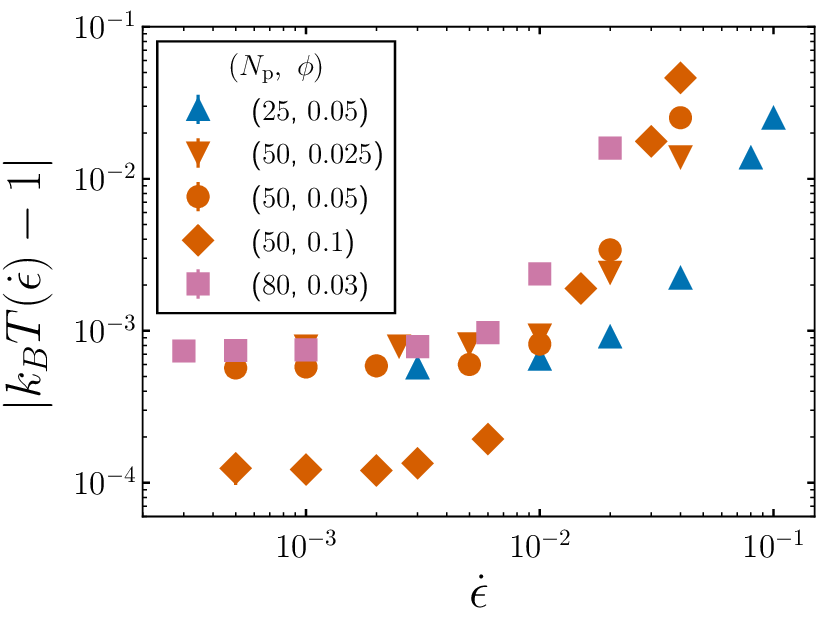} 
  \end{overpic}
  \caption{Relative temperature error $|k_BT(\dot{\epsilon})-1|$ as a function of the extension rate $\dot{\epsilon}$ for various $N_\mathrm{p}$ and $\phi$. The error bars denote the standard deviations from three independent simulations and are smaller than the symbol size.}

  \label{fig:temperature}
\end{figure}

We adopt the modified velocity Verlet method~\cite{Groot1997-je} for time integration, where the parameter $\lambda$ introduced in this scheme and the time step $\Delta t$ are set to $\lambda=0.65$ and $\Delta t=0.04$, respectively.
Figure~\ref{fig:temperature} shows the relative temperature error $|k_BT(\dot{\epsilon})-1|$ as a function of the extension rate $\dot{\epsilon}$, where $k_BT(\dot{\epsilon})$ is computed as $k_BT(\dot{\epsilon}) = \langle \bm{p}^2\rangle/3$ with $\langle\cdot \rangle $ denoting the ensemble average.
Since $|k_BT(\dot{\epsilon})-1|$ increases with $\dot{\epsilon}$, as reported for extensional-flow simulations of the Kremer--Grest model with the Langevin thermostat~\cite{Murashima2018-te}, we consider only extension rates satisfyiiing $|k_BT(\dot{\epsilon})-1|<{0.05}$.
The initial simulation box is a cube with dimensions $60\times 60\times 60$.
From a random initial configuration, equilibrium simulations are conducted for $12\,000$ time units, which is sufficiently long compared with the longest relaxation time of polymers considered.
After this initial equilibration, we apply uniaxial extensional flow to polymer solutions.
All DPD simulations are conducted with our in-house code.

\section{RESULTS AND DISCUSSION}

\begin{figure}
  \centering
  \begin{overpic}[width=0.75\linewidth]{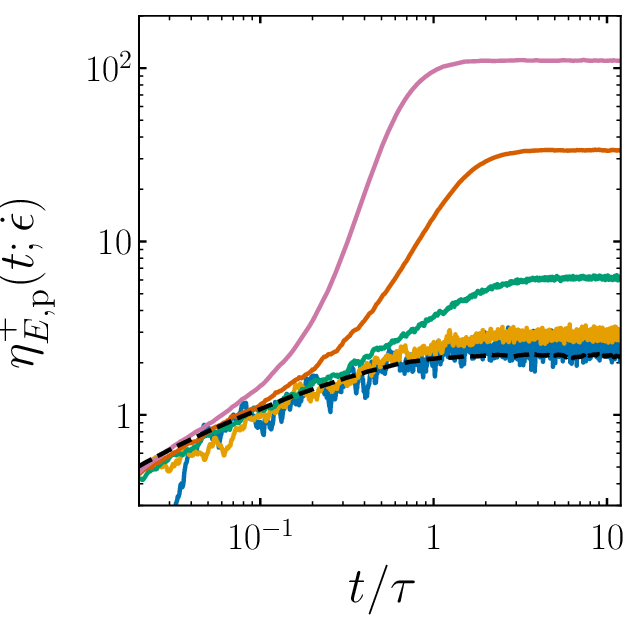} 
  \end{overpic}
  \caption{Polymer contribution $\eta_{E,\mathrm{p}}^+(t;\dot{\epsilon})$ to the extensional viscosity growth function as a function of the time $t$ normalized by the longest relaxation time $\tau$ of polymers for $N_\mathrm{p}=50$ and $\phi=0.1$. 
  From bottom to top, the curves correspond to $\mathrm{Wi}=0.48$, $0.80$, $1.6$, $3.2$, and $6.4$. The dashed line shows the LVE envelope $3\eta_{0,\mathrm{p}}^+(t)$.}
  \label{fig:vis_growth}
\end{figure}

We first investigate the extensional viscosity growth function of dilute polymer solutions under start-up uniaxial extensional flow.
Specifically, for $t>0$, we impose a uniaxial extensional flow on the systems that are initially at equilibrium.
The extensional viscosity growth function $\eta^+_E(t;\dot{\epsilon})$ is defined as 
\begin{equation}
  \eta_E^+(t;\dot{\epsilon}) = \frac{\sigma_{xx}^+(t;\dot{\epsilon})-\{\sigma_{yy}^+(t;\dot{\epsilon})+\sigma_{zz}^+(t;\dot{\epsilon})\}/2}{\dot{\epsilon}}, \label{eq:viscosity_growth}
\end{equation}
where $\sigma_{\alpha\beta}^+(t;\dot{\epsilon})$ is the stress growth function.
We employ the Irving--Kirkwood stress tensor~\cite{Irving1950-lc,Liu2015-yj} to calculate $\sigma_{\alpha\beta}^+(t;\dot{\epsilon})$:
\begin{equation}
  \sigma_{\alpha\beta}^+(t;\dot{\epsilon}) = -\frac{1}{V}\left(\sum_{i<j} r_{ij,\alpha}(t)F_{ij,\beta}(t)+\sum_{i} p_{i,\alpha}(t)p_{i,\beta}(t)\right), 
\end{equation}
where $V$ is the system volume, $\bm{F}_{ij}(t)$ is the force exerted on the $i$-th particle by the $j$-th particle, and $\bm{p}_i(t)$ is the peculiar momentum of the $i$-th particle.
We evaluate $\eta_E^+(\dot{t;\epsilon})$ by averaging the results from 20 independent simulations.
In addition, we apply a moving average with a window of $100\Delta t$ to reduce the noise.
Here, we focus on the polymer contribution $\eta_{E,\mathrm{p}}^{+}(t;\dot{\epsilon})=\eta_E^+(t;\dot{\epsilon})-\eta_{E,\mathrm{s}}$ to the extensional viscosity growth function, where $\eta_{E,\mathrm{s}}$ is the extensional viscosity of the solvent obtained from additional DPD simulations of a system consisting only of solvent particles at a certain $\dot{\epsilon}$.
Figure~\ref{fig:vis_growth} shows $\eta_{E,\mathrm{p}}^+(t;\dot{\epsilon})$ as a function of the time $t$ normalized by the longest relaxation time $\tau$ of polymers for $N_\mathrm{p}=50$ and $\phi=0.1$ at various Weissenberg numbers $\mathrm{Wi}=\tau\dot{\epsilon}$.
We evaluate $\tau$ by fitting the autocorrelation function $C(t)$ of the end-to-end vector to an exponential function $C(t) = C_0\exp(-t/\tau)$~\cite{Jiang2007-rf}. 
We also show the linear viscoelastic~(LVE) envelope $3\eta_{0,\mathrm{p}}^+(t)(=3\eta_{0}^+(t)-\eta_{E,\mathrm{s}})$, where $\eta_{0}^+(t)$ is computed from the linear relaxation modulus $G(t)$ as 
\begin{equation}
  \eta_{0}^+(t) = \int_{0}^t G(t^\prime)dt^\prime.
\end{equation}
We obtain $G(t)$ from the stress autocorrelation function at equilibrium using the Green-Kubo formula~\cite{evans_morriss_2008}.
We confirm that at $\mathrm{Wi}=0.48$, the relation $\eta_{E,\mathrm{p}}^+(t;\dot{\epsilon})\simeq 3\eta_{0,\mathrm{p}}^+(t)$ holds, indicating the linear viscoelastic behavior under weak uniaxial extensional flow.
In contrast, at $\mathrm{Wi}\geq 0.80$, $\eta_{E,\mathrm{p}}^+(t;\dot{\epsilon})$ exhibits strain hardening and then approaches a plateau, while it still follows the LVE envelope at short times.
This strain hardening at large $\mathrm{Wi}$ is consistent with observations in previous experiments and Brownian dynamics simulations of dilute polymer solutions~\cite{van-den-Brule1993-mg,Gupta2000-ad,Li2000-da,Anna2001-bn,Hsieh2003-oc,Larson2005-ef,Dinic2015-ec}.

\begin{figure}
  \centering
  \begin{overpic}[width=0.75\linewidth]{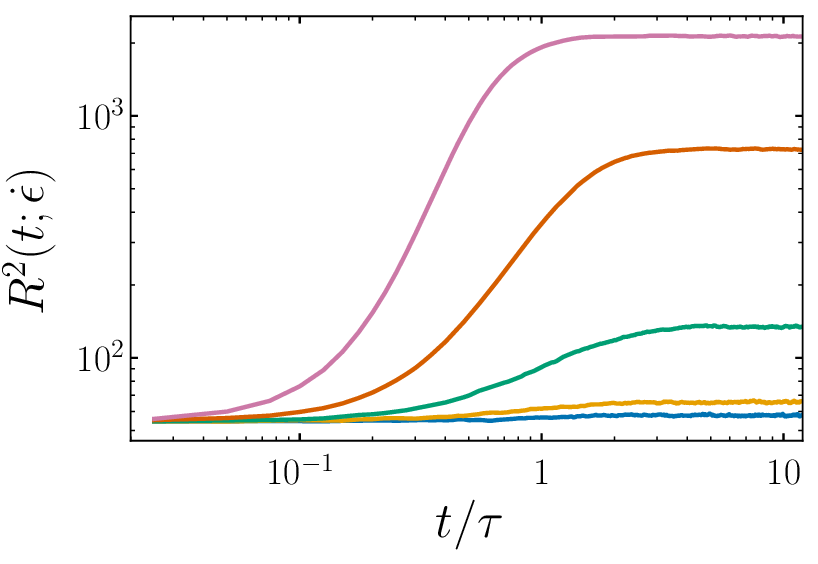} 
  \end{overpic}
  \caption{Mean-square end-to-end distance $R^2(t;\dot{\epsilon})$ as a function of $t/\tau$ for $N_\mathrm{p}=50$ and $\phi=0.1$. From bottom to top, the curves correspond to $\mathrm{Wi}=0.48$, $0.80$, $1.6$, $3.2$, and $6.4$.}
  \label{fig:endtoend_growth}
\end{figure}

To elucidate the physical mechanism behind the strain hardening, we examine the conformational changes of polymers under start-up uniaxial extensional flow.
We first focus on the end-to-end distance of polymers.
Figure~\ref{fig:endtoend_growth} shows the time evolution of the mean-square end-to-end distance $R^2(t;\dot{\epsilon})$ for the same values of $\mathrm{Wi}$ as in Fig.~\ref{fig:vis_growth}.
Here, $R^2(t;\dot{\epsilon})$ is defined as 
\begin{equation}
  R^2(t;\dot{\epsilon}) = \left\langle |\bm{r}_{N_\mathrm{p}}-\bm{r}_{1}|^2\right\rangle_{t,\dot{\epsilon}},
\end{equation}
where $\left\langle\cdot \right\rangle_{t,\dot{\epsilon}}$ denotes the average over all polymers in the system at time $t$ under a given $\dot{\epsilon}$.
Similar to $\eta_{E,\mathrm{p}}^+(t;\dot{\epsilon})$, $R^2(t;\dot{\epsilon})$ increases with $t$, and the growth becomes more pronounced as $\mathrm{Wi}$ increases.
This increasing tendency of $R^2(t;\dot{\epsilon})$ indicates the polymer stretching by strong uniaxial extensional flows, consistent with previous DPD simulations~\cite{Jana2025-sh}.
Such highly stretched states of polymers are specific to extensional flows, which are known as coil-stretch transition~\cite{De_Gennes1974-md}.

\begin{figure}
  \centering
        \centering
          \begin{overpic}[width=0.7\linewidth]{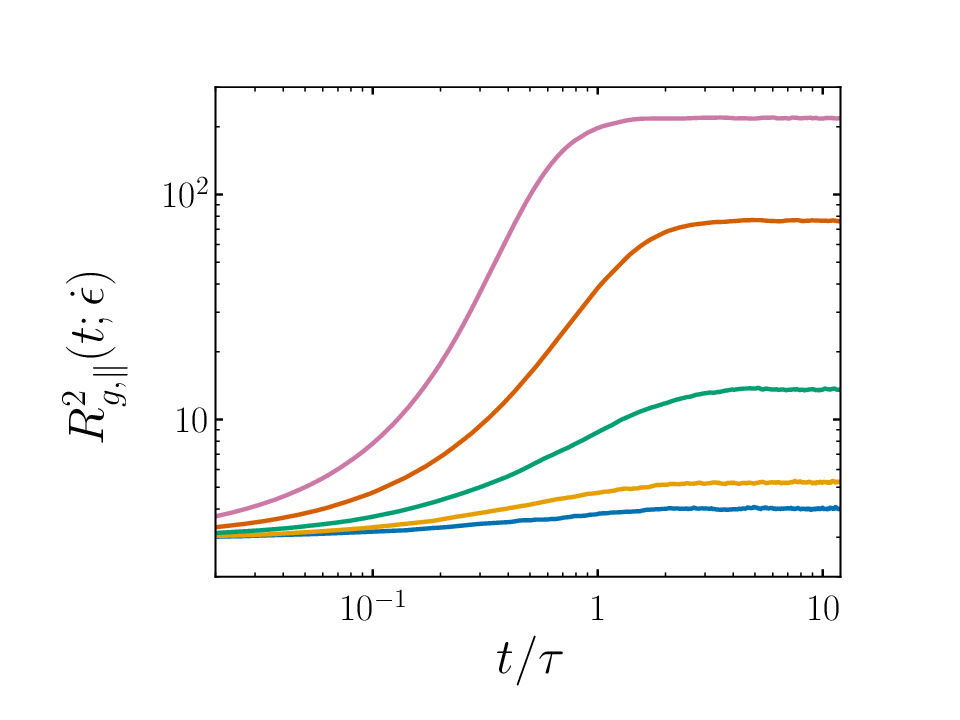}
              \linethickness{3pt}
        \put(8,65){(a)}

          \end{overpic}
          \vspace{0.2em}
        \centering
          \begin{overpic}[width=0.7\linewidth]{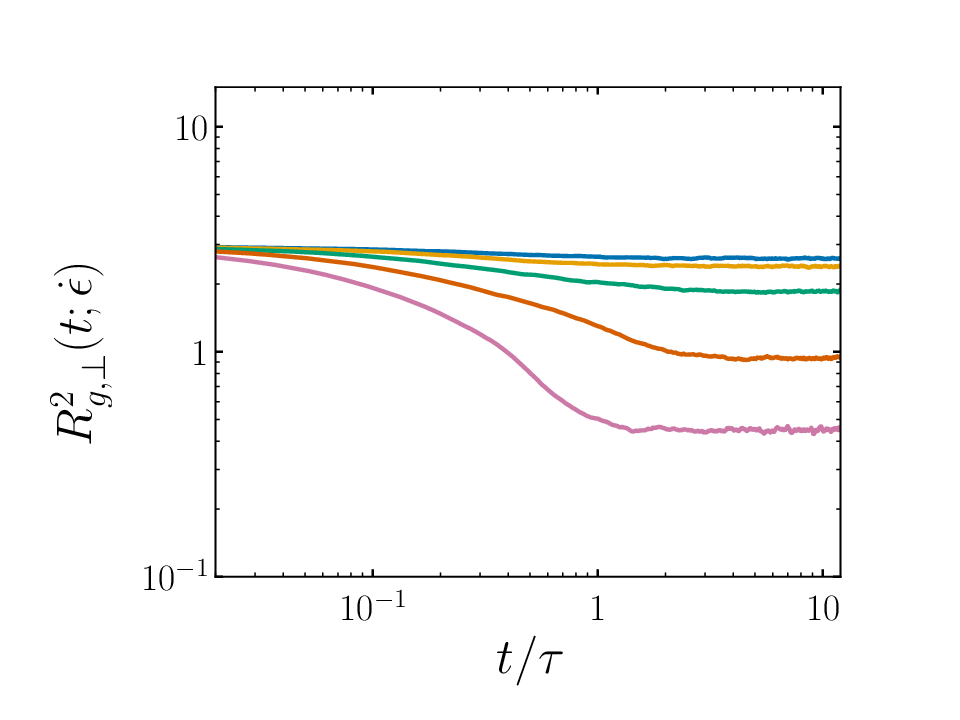}
              \linethickness{3pt}
        \put(8,65){(b)}

          \end{overpic}

    \caption{Gyration radius of polymers in (a) the parallel $R_{g,\parallel}(t;\dot{\epsilon})$ and (b) perpendicular $R_{g,\perp}(t;\dot{\epsilon})$ directions to the extensional direction as functions of $t/\tau$ for $N_\mathrm{p}=50$ and $\phi=0.1$. 
    From bottom to top in (a), and from top to bottom in (b), the curves correspond to $\mathrm{Wi}=0.48$, $0.80$, $1.6$, $3.2$, and $6.4$.}

      \label{fig:gyration_growth}
\end{figure}

Next, we investigate the gyration radius of polymers in each direction as another common metric for quantifying polymer conformation.
Considering that the $y$ and $z$ directions are symmetric, we introduce the gyration radii in the parallel ($R_{g,\parallel}(t;\dot{\epsilon})$) and perpendicular ($R_{g,\perp}(t;\dot{\epsilon})$) directions to the extensional direction~(i.e., the $x$ direction) as
\begin{equation}
  R_{g,\parallel}^2(t;\dot{\epsilon}) = \left\langle \frac{1}{N_\mathrm{p}}\sum_{i=1}^{N_{\mathrm{p}}} (r_{i,x}-r_{G,x})^2\right\rangle_{t,\dot{\epsilon}},
\end{equation}
\begin{equation}
  R_{g,\perp}^2(t;\dot{\epsilon}) = \frac{1}{2}\left\langle \frac{1}{N_\mathrm{p}}\sum_{i=1}^{N_{\mathrm{p}}} \{(r_{i,y}-r_{G,y})^2+(r_{i,z}-r_{G,z})^2\}\right\rangle_{t,\dot{\epsilon}},
\end{equation}
where $\bm{r}_G$ is the position of the center of mass of polymers. 
Figure~\ref{fig:gyration_growth} shows $R_{g,\parallel}^2(t;\dot{\epsilon})$ and $R_{g,\perp}^2(t;\dot{\epsilon})$ as functions of $t/\tau$ for the same values of $\mathrm{Wi}$ as in Fig.~\ref{fig:vis_growth}.
We observe that $R_{g,\parallel}^2(t;\dot{\epsilon})$ increases with $t$, and the growth becomes more significant for larger $\mathrm{Wi}$.
In contrast, $R_{g,\perp}^2(t;\dot{\epsilon})$ gradually decreases as $t$ increases.
The observed behavior of $R_{g,\parallel}^2(t;\dot{\epsilon})$ and $R_{g,\perp}^2(t;\dot{\epsilon})$ reflects the polymer stretching by uniaxial extensional flow, consistent with the trends in $R^2(t;\dot{\epsilon})$~(Fig.~\ref{fig:endtoend_growth}).
Thus, as is well known, the pronounced increase in $\eta_{E,\mathrm{p}}^+(t;\dot{\epsilon})$ can be qualitatively attributed to the polymer stretching in the extensional direction.
However, from a quantitative viewpoint, it remains unclear whether the end-to-end distance or the gyration radius is more directly related to $\eta_{E,\mathrm{p}}^+(t;\dot{\epsilon})$. 

\begin{figure}
  \centering
  \begin{overpic}[width=0.75\linewidth]{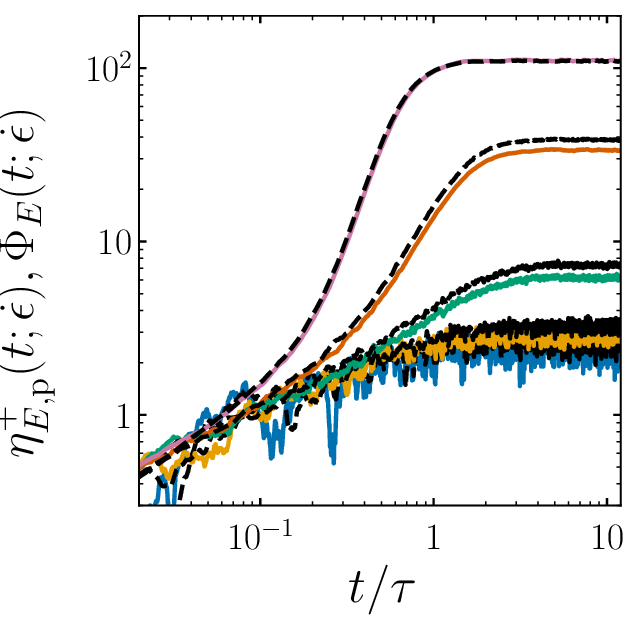} 
  \end{overpic}
  \caption{Polymer contribution $\eta_{E,\mathrm{p}}^+(t;\dot{\epsilon})$ to the extensional viscosity growth function~(solid lines) and $\Phi_E(t;\dot{\epsilon})$~(dashed lines) as functions of $t/\tau$ for $N_\mathrm{p}=50$ and $\phi=0.1$.
  From bottom to top, the curves correspond to $\mathrm{Wi}=0.48$, $0.80$, $1.6$, $3.2$, and $6.4$.}
  \label{fig:viscosity_polymer_contribution_growth}
\end{figure}

We then identify the quantitative relation between $\eta_{E,\mathrm{p}}^+(t;\dot{\epsilon})$ and polymer conformation.
For this purpose, we adopt the Rouse-type model, which is a single-chain model consisting of beads interacting via interaction potentials.
In the Rouse-type model, each bead follows the overdamped Langevin equation
\begin{equation}
    \frac{d\bm{r}_i(t)}{dt} = -\frac{1}{\zeta}\frac{\partial \mathcal{U}(\{\bm{r}_j(t)\})}{\partial \bm{r}_i(t)}+(\nabla \bm{u})^\mathsf{T}\cdot \bm{r}_i(t)+\sqrt{\frac{2k_BT}{\zeta}}\bm{w}_i(t),\label{eq:Rouse}
\end{equation}
where $\bm{r}_i(t)$ is the position of the $i$-th bead, $\zeta$ is the friction coefficient, $\mathcal{U}(\{\bm{r}_j(t)\})$ is the interaction potential for a single polymer chain, and $\bm{w}_i(t)$ is the random variable that satisfies
\begin{align}
    &\langle \bm{w}_i(t)\rangle =0,\\
    &\langle \bm{w}_i(t)\bm{w}_j(s)\rangle =\delta_{ij}\delta(t-s)\bm{I},
\end{align}
where $\delta_{ij}$ is the Kronecker delta, $\delta(t)$ is the delta function, and $\bm{I}$ is the indentity tensor.
Eq.~\eqref{eq:Rouse} can be viewed as a coarse-grained, approximate equation of motion for a single tagged polymer after eliminating the degrees of freedom of the surrounding solvent and other chains, although hydrodynamic interactions present in DPD~\cite{Jiang2007-rf} are not explicitly included.
We will discuss the role of the hydrodynamic interactions in our analysis based on the Rouse-type model at the end of this section.
Uneyama~\cite{uneyama2025radius} theoretically derived a relation connecting $\eta_{E,\mathrm{p}}^+(t;\dot{\epsilon})$ to $R_{g,\parallel}^2(t;\dot{\epsilon})$ and $R_{g,\perp}^2(t;\dot{\epsilon})$ for the Rouse-type model:
\begin{equation}
    \eta_{E,\mathrm{p}}^+(t;\dot{\epsilon}) = \rho_\mathrm{p}\zeta \left[R_{g,\parallel}^2(t;\dot{\epsilon})+\frac{1}{2}R_{g,\perp}^2(t;\dot{\epsilon})-\frac{1}{2\dot{\epsilon}} \frac{d}{dt}\left\{R_{g,\parallel}^2(t;\dot{\epsilon})-R_{g,\perp}^2(t;\dot{\epsilon})\right\}\right]\label{eq:Rouse_vis_growth},
\end{equation}
where $\rho_\mathrm{p}$ is the number density of polymer beads.
We denote the right-hand side of Eq.~\eqref{eq:Rouse_vis_growth} by $\Phi_E(t;\dot{\epsilon})$, which is computed from the gyration radius of polymers under uniaxial extensional flow.
Figure~\ref{fig:viscosity_polymer_contribution_growth} shows the time evolution of $\eta_{E,\mathrm{p}}^+(t;\dot{\epsilon})$ and $\Phi_E(t;\dot{\epsilon})$.
Since $\zeta$ is an a priori unknown parameter, we estimate $\zeta$ using the relation for the zero-shear viscosity $\eta_0$ in the Rouse-type model~\cite{uneyama2025radius}:
\begin{equation}
  \eta_0 = \frac{\rho_\mathrm{p}\zeta }{6}R_{g,\mathrm{eq}}^2,\label{eq:zeroshear}
\end{equation}
where $R_{g,\mathrm{eq}}$ is the gyration radius at equilibrium and $\eta_0=\int_0^\infty G(t^\prime)dt^\prime$ is computed using the time decomposition method~\cite{Zhang2015-oi,Panoukidou2021-gz}.
Accordingly, we use a single value of $\zeta$ for all $\mathrm{Wi}$.
We find that $\Phi_E(t;\dot{\epsilon})$ agrees well with $\eta_{E,\mathrm{p}}^+(t;\dot{\epsilon})$ and quantitatively captures the strain-hardening behavior.
Thus, the Rouse-type model enables us to quantitatively describe the extensional viscosity growth function using the gyration radius of polymers.
Although $\Phi_E(t;\dot{\epsilon})\simeq \eta_{E,\mathrm{p}}^+(t;\dot{\epsilon})$ holds at short times irrespective of $\mathrm{Wi}$, a small discrepancy between $\Phi_E(t;\dot{\epsilon})$ and $\eta_{E,\mathrm{p}}^+(t;\dot{\epsilon})$ is observed at long times for $0.80\leq \mathrm{Wi}\leq 3.2$.
This deviation indicates that strong extensional flow can modulate $\zeta$.
We will discuss the details of $\zeta$ later.

In more detail, we decompose $\Phi_E(t;\dot{\epsilon})$ into three components:
\begin{equation}
  \Phi_E(t;\dot{\epsilon})=\Phi_{E,\parallel}(t;\dot{\epsilon})+\Phi_{E,\perp}(t;\dot{\epsilon})-\Phi_{E,\Delta}(t;\dot{\epsilon}),
\end{equation}
where $\Phi_{E,\parallel}(t;\dot{\epsilon})$, $\Phi_{E,\perp}(t;\dot{\epsilon})$, and $\Phi_{E,\Delta}(t;\dot{\epsilon})$ are defined as 
\begin{equation}
  \Phi_{E,\parallel}(t;\dot{\epsilon}) = \rho_\mathrm{p}\zeta R_{g,\parallel}^2(t;\dot{\epsilon}), 
\end{equation}
\begin{equation}
  \Phi_{E,\perp}(t;\dot{\epsilon})= \frac{1}{2}\rho_\mathrm{p}\zeta R_{g,\perp}^2(t;\dot{\epsilon}),
\end{equation}
\begin{equation}
   \Phi_{E,\Delta}(t;\dot{\epsilon})= \frac{1}{2\dot{\epsilon}}\rho_\mathrm{p}\zeta \frac{d}{dt}\left\{R_{g,\parallel}^2(t;\dot{\epsilon})-R_{g,\perp}^2(t;\dot{\epsilon})\right\}.
\end{equation}
Figure~\ref{fig:viscosity_polymer_contribution_growth}(b) shows $\Phi_E(t;\dot{\epsilon})$, $\Phi_{E,\parallel}(t;\dot{\epsilon})$, $\Phi_{E,\perp}(t;\dot{\epsilon})$, and $\Phi_{E,\Delta}(t;\dot{\epsilon})$ at $\mathrm{Wi}=0.80$ and $6.4$.
We observe that although $\Phi_{E,\parallel}(t;\dot{\epsilon})$ is the dominant component of $\Phi_E(t;\dot{\epsilon})$, the contributions from $\Phi_{E,\perp}(t;\dot{\epsilon})$ and $\Phi_{E,\Delta}(t;\dot{\epsilon})$ are also nonnegligible, especially in the transient regime.
This decomposition analysis indicates that a quantitative description of the transient behavior of $\eta_{E,\mathrm{p}}^+(t;\dot{\epsilon})$ requires not only the instantaneous gyration radius in each direction but also its time derivative.
\begin{figure}
  \centering
      \begin{tabular}{cc}
      \begin{minipage}{0.5\hsize}
          \begin{overpic}[width=0.9\linewidth]{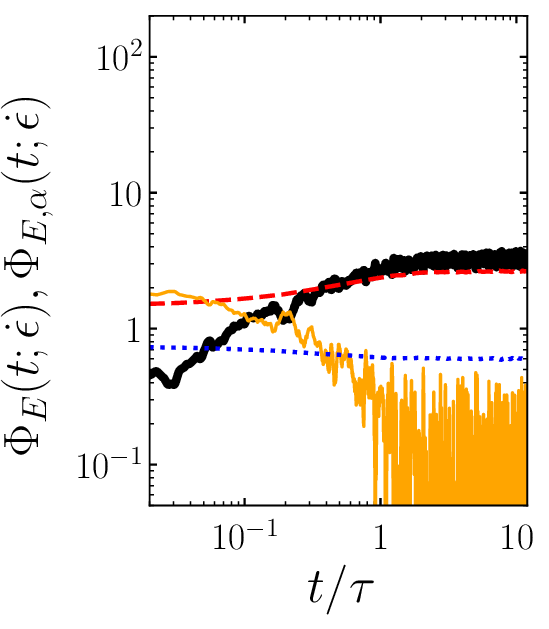}
              \linethickness{3pt}
        \put(5,95){(a)}

          \end{overpic}
      \end{minipage}
      \begin{minipage}{0.5\hsize}
          \begin{overpic}[width=0.9\linewidth]{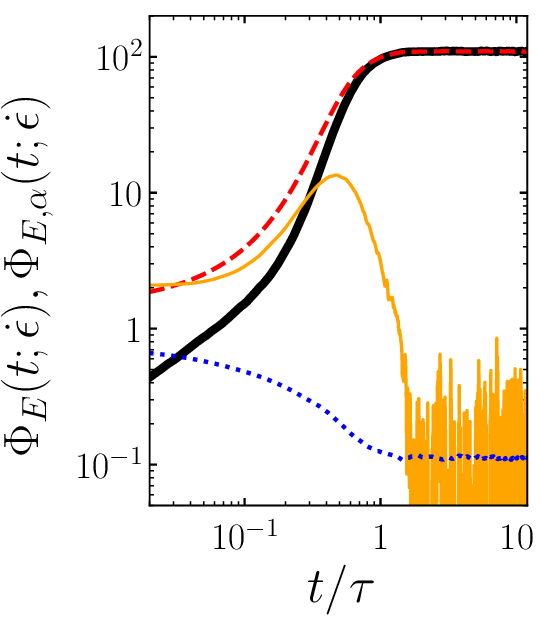}
              \linethickness{3pt}
        \put(5,95){(b)}

          \end{overpic}
      \end{minipage}
      \end{tabular}
      \caption{$\Phi_E(t;\dot{\epsilon})$~(thick solid line), $\Phi_{E,\parallel}(t;\dot{\epsilon})$~(dashed line), $\Phi_{E,\perp}(t;\dot{\epsilon})$~(dotted line), and $\Phi_{E,\Delta}(t;\dot{\epsilon})$~(thin solid line) as functions of $t/\tau$ at (a) $\mathrm{Wi}=0.80$ and (b) $6.4$ for $N_\mathrm{p}=50$ and $\phi=0.1$.}
      \label{fig:gyration_decomp}
\end{figure}

\begin{figure}
  \centering
  \begin{overpic}[width=0.75\linewidth]{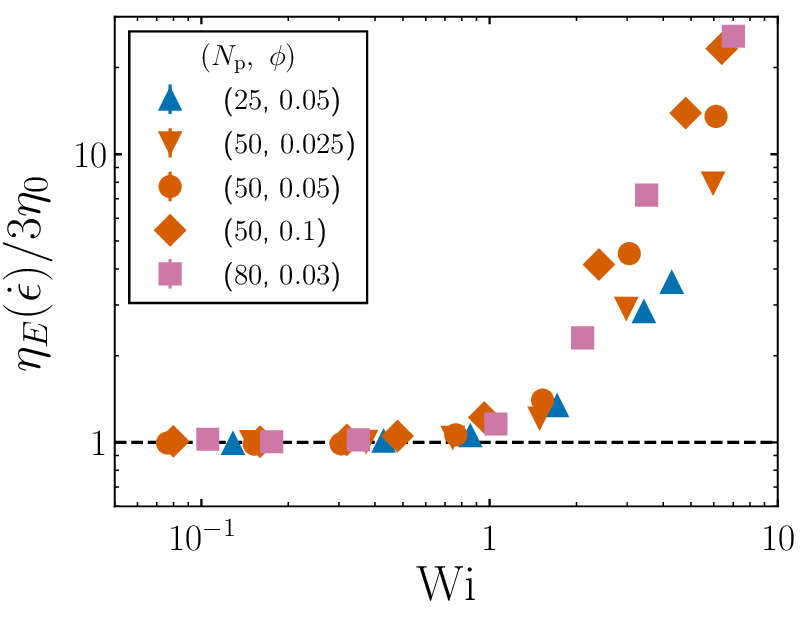} 
  \end{overpic}
  \caption{Steady-state extensional viscosity $\eta_E(\dot{\epsilon})$ normalized by three times the zero shear viscosity $3\eta_0$ as a function of the Weissenberg number $\mathrm{Wi}$. 
  The dashed line indicates the Newtonian limit $\eta_E(\dot{\epsilon})/3\eta_0=1$. The error bars denote the standard deviations from three independent simulations and are smaller than the symbol size.}
  \label{fig:steady_vis}
\end{figure}

Next, we examine the extension-rate dependence of the steady-state extensional viscosity $\eta_E(\dot{\epsilon})$ for various $N_\mathrm{p}$ and $\phi$.
We obtain $\eta_E(\dot{\epsilon})$ by averaging values of $\eta_{E}^+(t;\dot{\epsilon})$ at steady states.
Figure~\ref{fig:steady_vis} shows $\eta_E(\dot{\epsilon})$ normalized by three times the zero shear viscosity $3\eta_0$ as a function of $\mathrm{Wi}=\tau\dot{\epsilon}$.
For $\mathrm{Wi}\lesssim 1$, $\eta_E(\dot{\epsilon})\approx 3\eta_0$, consistent with the linear viscoelastic behavior observed in Fig.~\ref{fig:vis_growth}.
For $\mathrm{Wi}\gtrsim 1$, $\eta_E(\dot{\epsilon})$ increases significantly with $\mathrm{Wi}$.
Similar behavior has been observed in Brownian dynamics simulations of polymer models~\cite{van-den-Brule1993-mg,Doyle1997-tq,Herrchen1997-fb,Neelov2002-cp,Stoltz2006-ht,Prabhakar2017-no} and in experiments on dilute polymer solutions~\cite{Ng1996-ed,Kim2019-xr,Haward2023b}.
One may argue that the onset of the increase in $\eta_E(\dot{\epsilon})$ should occur around $\mathrm{Wi}= 1/2$.
Note that we define the Weissenberg number $\mathrm{Wi}=\tau\dot{\epsilon}$ using the longest relaxation time $\tau$ determined from polymer conformation, rather than from stress.
Given that the former is twice the latter for the Rouse model~\cite{rubinstein2003polymer}, the observed onset near $\mathrm{Wi}=1$ is reasonable.
Our simulations are limited to $\mathrm{Wi}\lesssim 10$ due to the pronounced temperature increase~(Fig.~\ref{fig:temperature}).
Consequently, $\eta_E(\dot{\epsilon})$ does not reach a plateau within the range of $\mathrm{Wi}$ considered.

\begin{figure}
  \centering
  \begin{overpic}[width=0.8\linewidth]{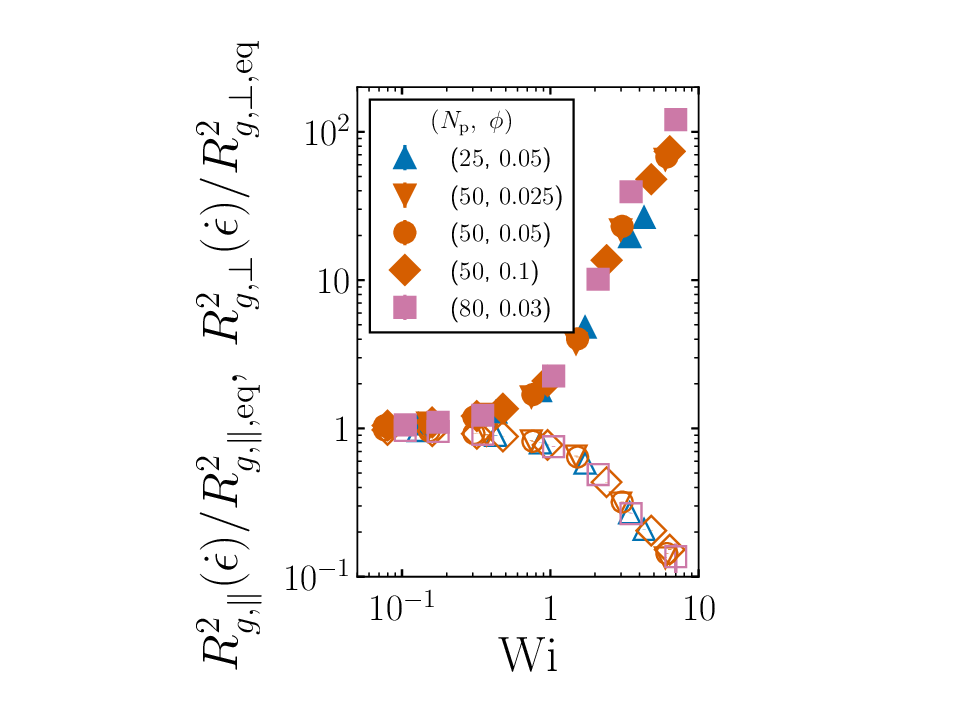} 
  \end{overpic}
  \caption{Normalized gyration radii of polymers in the parallel $R_{g,\parallel}(\dot{\epsilon})/R_{g,\parallel,\mathrm{eq}}$~(filled symbols) and perpendicular $R_{g,\perp}(\dot{\epsilon})/R_{g,\perp,\mathrm{eq}}$~(open symbols) directions to the extensional direction in the steady state as functions of the Weissenberg number $\mathrm{Wi}$.
  The error bars denote the standard deviations from three independent simulations and are smaller than the symbol size.}

  \label{fig:gyration}
\end{figure}

To explore the origin of the increase in $\eta_E(\dot{\epsilon})$, we investigate the conformational changes of polymers at steady states.
Since we have demonstrated that the gyration radius is relevant to the extensional viscosity in Fig.~\ref{fig:viscosity_polymer_contribution_growth}, we focus on the gyration radii in the parallel ($R_{g,\parallel}(\dot{\epsilon})$) and perpendicular ($R_{g,\perp}(\dot{\epsilon})$) directions to the extensional direction.
Here, we evaluate $R_{g,\parallel}^2(\dot{\epsilon})$ and $R_{g,\perp}^2(\dot{\epsilon})$ by averaging $R_{g,\parallel}^2(t;\dot{\epsilon})$ and $R_{g,\perp}^2(t;\dot{\epsilon})$ at steady states, respectively.
Figure~\ref{fig:gyration} shows $R_{g,\parallel}^2(\dot{\epsilon})$ and $R_{g,\perp}^2(\dot{\epsilon})$ normalized by their equilibrium values $R_{g,\parallel,\mathrm{eq}}^2$ and $R_{g,\perp,\mathrm{eq}}^2$ as functions of $\mathrm{Wi}$.
We observe that $R_{g,\parallel}^2(\dot{\epsilon})$ increases significantly for $\mathrm{Wi}\gtrsim 1$, while $R_{g,\perp}^2(\dot{\epsilon})$ monotonically decreases with increasing $\mathrm{Wi}$.
These results indicate that for $\mathrm{Wi}\gtrsim 1$, polymers are aligned and stretched in the extensional direction, which qualitatively explains the increase in $\eta_E(\dot{\epsilon})$~(Fig.~\ref{fig:steady_vis}).

\begin{figure}
  \centering
  \begin{overpic}[width=0.9\linewidth]{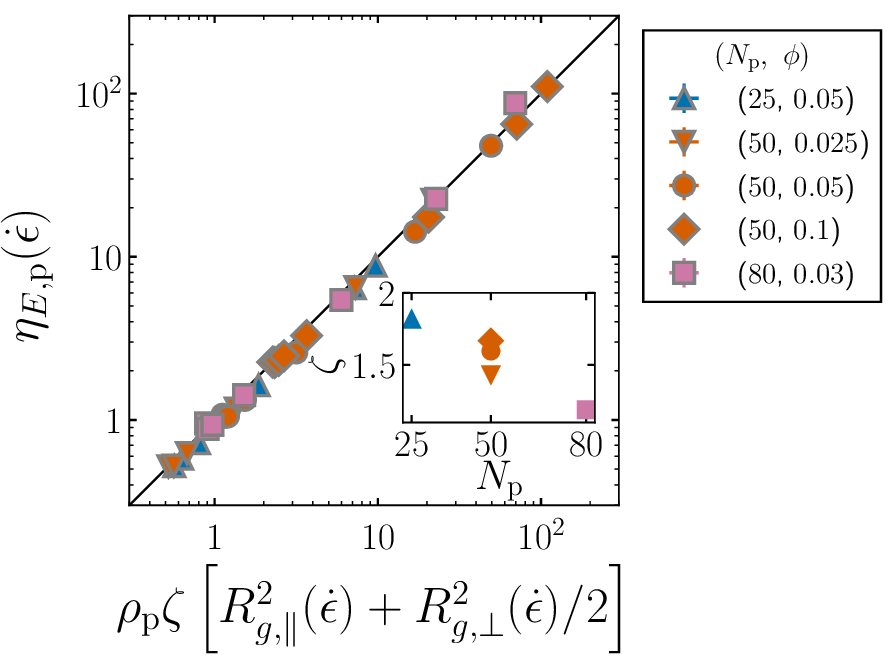} 
  \end{overpic}
  \caption{Polymer contribution $\eta_{E,\mathrm{p}}(\dot{\epsilon})$ to the steady-state extensional viscosity $\eta_E(\dot{\epsilon})$ as a function of $\rho_\mathrm{p} \zeta [R_{g,\parallel}^2(\dot{\epsilon})+R_{g,\perp}^2(\dot{\epsilon})/2]$ for different values of $N_\mathrm{p}$, $\phi$, and $\dot{\epsilon}$. The black line indicates $\eta_{E,\mathrm{p}}(\dot{\epsilon}) = \rho_\mathrm{p}\zeta [R_{g,\parallel}^2(\dot{\epsilon})+R_{g,\perp}^2(\dot{\epsilon})/2]$. The error bars denote the standard deviations from three independent simulations and are smaller than the symbol size. The inset shows $\zeta$ obtained with Eq.~\eqref{eq:zeroshear} as a function of $N_\mathrm{p}$.}
  \label{fig:steady_vis_gyration}
\end{figure}

\begin{figure}
  \centering
  \begin{overpic}[width=0.9\linewidth]{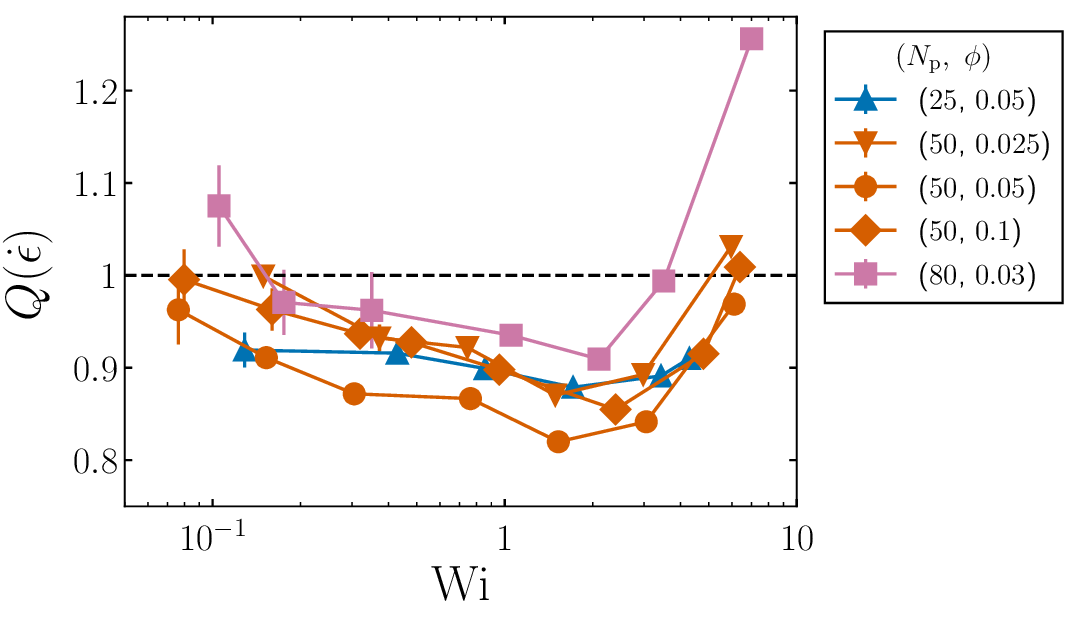} 
  \end{overpic}
  \caption{Ratio $Q(\dot{\epsilon})$ of $\eta_{E,\mathrm{p}}(\dot{\epsilon})$ to $\rho_\mathrm{p} \zeta [R_{g,\parallel}^2(\dot{\epsilon})+R_{g,\perp}^2(\dot{\epsilon})/2]$ as a function of the Weissenberg number $\mathrm{Wi}$ for different values of $N_\mathrm{p}$ and $\phi$. The black dashed line indicates $Q(\dot{\epsilon})=1$. The error bars denote the standard deviations from three independent simulations.}
  \label{fig:steady_vis_gyration_ratio}
\end{figure}

We aim to quantitatively connect $\eta_E(\dot{\epsilon})$ with the gyration radius of polymers using the Rouse-type model.
At the steady state, Eq.~\eqref{eq:Rouse_vis_growth} reduces to
\begin{equation}
    \eta_{E,\mathrm{p}}(\dot{\epsilon}) = \rho_\mathrm{p}\zeta \left[R_{g,\parallel}^2(\dot{\epsilon})+\frac{1}{2}R_{g,\perp}^2(\dot{\epsilon})\right]\label{eq:Rouse_vis_steady},
\end{equation}
where $\eta_{E,\mathrm{p}}(\dot{\epsilon})=\eta_E(\dot{\epsilon})-\eta_{E,\mathrm{s}}$ denotes the polymer contribution to the extensional viscosity.
On the basis of Eq.~\eqref{eq:Rouse_vis_steady}, Fig.~\ref{fig:steady_vis_gyration} shows $\eta_{E,\mathrm{p}}(\dot{\epsilon})$ as a function of $\rho_\mathrm{p} \zeta \left[R_{g,\parallel}^2(\dot{\epsilon})+\frac{1}{2}R_{g,\perp}^2(\dot{\epsilon})\right]$ by replotting the data from Figs.~\ref{fig:steady_vis} and \ref{fig:gyration}. 
The inset of Fig.~\ref{fig:steady_vis_gyration} shows the values of $\zeta$ obtained from Eq.~\eqref{eq:zeroshear} for each $N_\mathrm{p}$ and $\phi$.
We observe that $\eta_{E,\mathrm{p}}(\dot{\epsilon})$ almost collapses onto the line $\eta_{E,\mathrm{p}}(\dot{\epsilon})= \rho_\mathrm{p} \zeta \left[R_{g,\parallel}^2(\dot{\epsilon})+\frac{1}{2}R_{g,\perp}^2(\dot{\epsilon})\right]$, including the parameter sets already reported in our previous work~\cite{Koide2025-zr}.
Thus, we demonstrate that the increase in $\eta_E(\dot{\epsilon})$ of dilute polymer solutions~(Fig.~\ref{fig:steady_vis}) can be quantitatively explained by flow-induced changes in the polymer gyration radii in the directions parallel $R_{g,\parallel}(\dot{\epsilon})$ and perpendicular $R_{g,\perp}(\dot{\epsilon})$ to the extensional direction.
For a more comprehensive evaluation of the validity of our approach, Fig.~\ref{fig:steady_vis_gyration_ratio} shows the ratio $Q(\dot{\epsilon})=\eta_{E,\mathrm{p}}(\dot{\epsilon})/\{\rho_\mathrm{p} \zeta [R_{g,\parallel}^2(\dot{\epsilon})+R_{g,\perp}^2(\dot{\epsilon})/2]\}$ of the left- and right-hand sides of Eq.~\eqref{eq:Rouse_vis_steady} as a function of $\mathrm{Wi}$.
Although $Q(\dot{\epsilon})$ stays close to unity, $Q(\dot{\epsilon})$ exhibits a slight nonmonotonic variation with respect to $\mathrm{Wi}$, consistent with Fig.~\ref{fig:viscosity_polymer_contribution_growth}.
This deviation can be attributed to the breakdown of the assumption that $\zeta$ is independent of $\mathrm{Wi}$.
To obtain a more complete description of $\eta_{E,\mathrm{p}}(\dot{\epsilon})$, investigating $\dot{\epsilon}$ dependence of the effective friction coefficient would be an important direction for future work.


\begin{figure}
  \centering
  \begin{overpic}[width=0.9\linewidth]{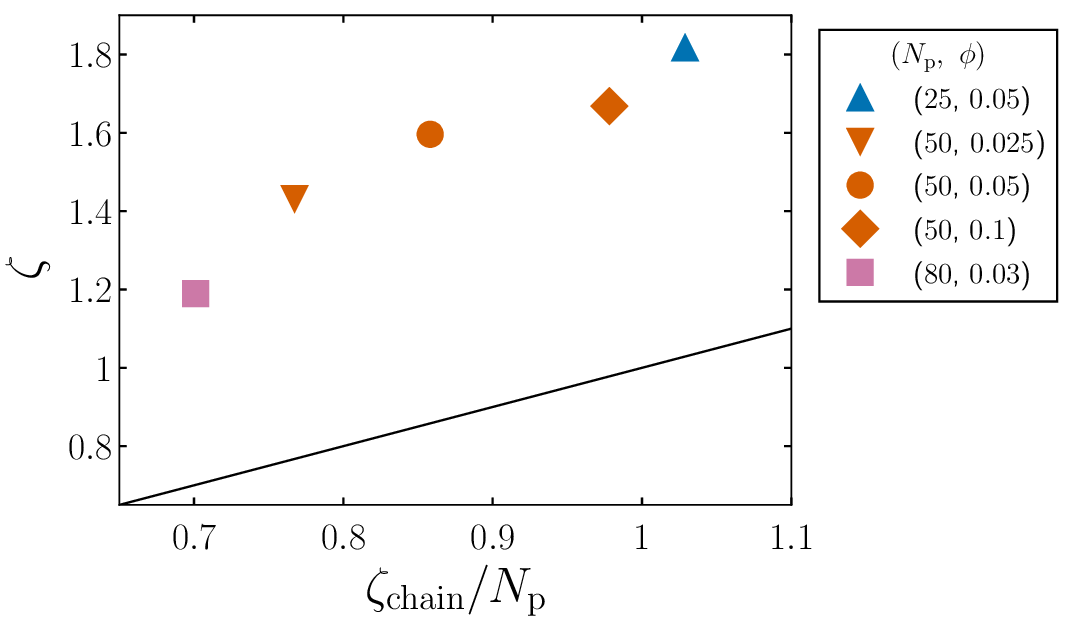} 
  \end{overpic}
  \caption{Friction coefficient $\zeta$ obtained from the Rouse-type model fit as a function of the chain friction coefficient $\zeta_\mathrm{chain}$ normalized by $N_\mathrm{p}$. The solid line indicates $\zeta=\zeta_\mathrm{chain}/N_\mathrm{p}$.}
  \label{fig:zeta}
\end{figure}

Before closing this section, we briefly discuss the physical origin of $\zeta$ in our DPD simulations.
We evaluate $\zeta$ from equilibrium simulations via Eq.~\eqref{eq:zeroshear} for each $N_\mathrm{p}$ and $\phi$.
The inset of Fig.~\ref{fig:steady_vis_gyration} demonstrates that $\zeta$ depends weakly on $N_\mathrm{p}$ and $\phi$.
To identify what influences $\zeta$, we compare the chain friction coefficient $N_\mathrm{p}{\zeta}$ obtained from the Rouse-type model fit with the chain friction coefficient $\zeta_\mathrm{chain}=k_BT/D_G$ obtained from the diffusion coefficient $D_G$ of the polymer center of mass.
Here, we evaluate $D_G$ from the mean-square displacement of the polymer center of mass.
Since we focus on $N_\mathrm{p}$ dependence of $\zeta$, Fig.~\ref{fig:zeta} shows the relation between $N_\mathrm{p}{\zeta}$ and $\zeta_\mathrm{chain}$ after dividing both quantities by $N_\mathrm{p}$.
Although ${\zeta}$ does not coincide exactly with $\zeta_\mathrm{chain}/N_\mathrm{p}$, they are positively correlated, indicating that $\zeta$ reflects the friction experienced by polymer chains.
Because the hydrodynamic interactions affect $N_\mathrm{p}$ dependence of $D_G$ in DPD simulations~\cite{Jiang2007-rf}, we infer that the hydrodynamic interaction effects, neglected in the Rouse-type model, are effectively absorbed into $\zeta$.
Thus, as in the present study, $\zeta$ should be evaluated for each $N_\mathrm{p}$ and $\phi$.
We also observe a slight breakdown of the approximation that $\zeta$ is independent of $\dot{\epsilon}$~(Fig.~\ref{fig:steady_vis_gyration_ratio}).
This is likely to reflect the variation in the effective friction due to the conformational changes under extensional flow~\cite{Hinch1977-lj,Fuller1981-mm,Magda1988-ha,Larson2005-ef,Prabhakar2016-ba,Prabhakar2017-no}.
However, since careful treatment of temperature increases is required for analyses at large $\mathrm{Wi}$, a detailed study of $\mathrm{Wi}$ dependence of $\zeta$ is left for future work.

\section{CONCLUSIONS}
We have revealed the quantitative relation between the extensional viscosity and the gyration radius in dilute polymer solutions, including both transient and steady states, by using dissipative particle dynamics simulations.
The extensional viscosity growth function $\eta_E^+(t;\dot{\epsilon})$ of dilute polymer solutions exhibits significant strain hardening for $\mathrm{Wi}\gtrsim 1$~(Fig.~\ref{fig:vis_growth}).
In this regime, polymers are highly stretched by uniaxial extensional flow~(Figs.~\ref{fig:endtoend_growth} and \ref{fig:gyration_growth}).
To quantitatively relate $\eta_E^+(t;\dot{\epsilon})$ to polymer conformation, we have applied a modeling approach based on the Rouse-type model.
Our analyses have revealed that the conformational change of polymers affects the polymer contribution $\eta_{E,\mathrm{p}}^+(t;\dot{\epsilon})$ to $\eta_E^+(t;\dot{\epsilon})$ through the instantaneous gyration radii in the parallel ($R_{g,\parallel}(t;\dot{\epsilon})$) and perpendicular ($R_{g,\perp}(t;\dot{\epsilon})$) directions to the extensional direction and their time derivatives~(Figs.~\ref{fig:viscosity_polymer_contribution_growth} and \ref{fig:gyration_decomp}).
We have also evaluated $\mathrm{Wi}$ dependence of the steady-state extensional viscosity $\eta_E(\dot{\epsilon})$~(Fig.~\ref{fig:steady_vis}).
The Rouse-type model has provided a unified description of the polymer contribution $\eta_{E,\mathrm{p}}(\dot{\epsilon})$ to $\eta_E(\dot{\epsilon})$ for various $N_\mathrm{p}$, $\phi$, and $\dot{\epsilon}$ in terms of the number density $\rho_\mathrm{p}$ of polymer particles, the effective friction coefficient $\zeta$, and the gyration radii in the parallel ($R_{g,\parallel}(\dot{\epsilon})$) and perpendicular ($R_{g,\perp}(\dot{\epsilon})$) directions to the extensional direction~(Fig.~\ref{fig:steady_vis_gyration}).

\section*{ACKNOWLEDGEMENTS}
This work was supported by JSPS Grants-in-Aid for Scientific Research (24KJ0109) and JST, ACT-X (JPMJAX24D5). 
The DPD simulations were mainly conducted using the HPCI Research Projects on supercomputer ``Flow'' at Information Technology Center, Nagoya University.


\begin{thebibliography}{10}

\bibitem{White2008-gb}
White CM, Mungal MG, \emph{Annu Rev Fluid Mech}, \textbf{40}, 235 (2008).

\bibitem{Steinberg2021-lz}
Steinberg V, \emph{Annu Rev Fluid Mech}, \textbf{53}, 27 (2021).

\bibitem{lumley1969drag}
Lumley JL, \emph{Annu Rev Fluid Mech}, \textbf{1}, 367 (1969).

\bibitem{Dimitropoulos1998-tl}
Dimitropoulos CD, Sureshkumar R, Beris AN, \emph{J Non-Newtonian Fluid Mech},
  \textbf{79}, 433 (1998).

\bibitem{Escudier1999-la}
Escudier MP, Presti F, Smith S, \emph{J Non-Newtonian Fluid Mech}, \textbf{81},
  197 (1999).

\bibitem{Serafini2022-ot}
Serafini F, Battista F, Gualtieri P, Casciola CM, \emph{Phys Rev Lett},
  \textbf{129}, 104502 (2022).

\bibitem{McKinley2000-ky}
McKinley GH, Tripathi A, \emph{J Rheol}, \textbf{44}, 653 (2000).

\bibitem{Anna2001-bn}
Anna SL, McKinley GH, \emph{J Rheol}, \textbf{45}, 115 (2001).

\bibitem{Dinic2015-ec}
Dinic J, Zhang Y, Jimenez LN, Sharma V, \emph{ACS Macro Lett}, \textbf{4}, 804
  (2015).

\bibitem{Clasen2006-rc}
Clasen C, Plog JP, Kulicke WM, Owens M, Macosko C, Scriven LE, Verani M,
  McKinley GH, \emph{J Rheol}, \textbf{50}, 849 (2006).

\bibitem{Tirtaatmadja2006-pa}
Tirtaatmadja V, McKinley GH, Cooper-White JJ, \emph{Phys Fluids}, \textbf{18},
  043101 (2006).

\bibitem{Matsuda2024-av}
Matsuda T, Sugiura R, Muto M, Tamano S, \emph{Phys Fluids}, \textbf{36}, 123120
  (2024).

\bibitem{Calabrese2024-dw}
Calabrese V, Shen AQ, Haward SJ, \emph{Macromolecules}, \textbf{57}, 9668
  (2024).

\bibitem{Calabrese2025-vc}
Calabrese V, Shen AQ, Haward SJ, \emph{Phys Rev X}, \textbf{15}, 021025 (2025).

\bibitem{Gaillard2024-ds}
Gaillard A, Herrada MA, Deblais A, Eggers J, Bonn D, \emph{Phys Rev Fluids},
  \textbf{9}, 073302 (2024).

\bibitem{Wang2025-df}
Wang HY, Prabhakar R, Prakash JR, Larson RG, \emph{J Rheol}, \textbf{69}, 641
  (2025).

\bibitem{Wu2020-vd}
Wu S, Mohammadigoushki H, \emph{Phys Rev Fluids}, \textbf{5}, 053303 (2020).

\bibitem{Haward2023a}
Haward SJ, Pimenta F, Varchanis S, Carlson DW, Toda-Peters K, Alves MA, Shen
  AQ, \emph{J Rheol}, \textbf{67}, 995 (2023).

\bibitem{Haward2023b}
Haward SJ, Varchanis S, McKinley GH, Alves MA, Shen AQ, \emph{J Rheol},
  \textbf{67}, 1011 (2023).

\bibitem{van-den-Brule1993-mg}
van~den Brule BHAA, \emph{J Non-Newtonian Fluid Mech}, \textbf{47}, 357 (1993).

\bibitem{Doyle1997-tq}
Doyle PS, Shaqfeh ESG, Gast AP, \emph{J Fluid Mech}, \textbf{334}, 251 (1997).

\bibitem{Herrchen1997-fb}
Herrchen M, ^^c3^^96ttinger HC, \emph{J Non-Newtonian Fluid Mech}, \textbf{68},
  17 (1997).

\bibitem{Li2000-da}
Li L, Larson RG, Sridhar T, \emph{J Rheol}, \textbf{44}, 291 (2000).

\bibitem{Somasi2002-ot}
Somasi M, Khomami B, Woo NJ, Hur JS, Shaqfeh ESG, \emph{J Non-Newtonian Fluid
  Mech}, \textbf{108}, 227 (2002).

\bibitem{Hsieh2003-oc}
Hsieh CC, Li L, Larson RG, \emph{J Non-Newtonian Fluid Mech}, \textbf{113}, 147
  (2003).

\bibitem{Larson2005-ef}
Larson RG, \emph{J Rheol}, \textbf{49}, 1 (2005).

\bibitem{De_Gennes1974-md}
De~Gennes PG, \emph{J Chem Phys}, \textbf{60}, 5030 (1974).

\bibitem{Dobson2014-kr}
Dobson M, \emph{J Chem Phys}, \textbf{141}, 184103 (2014).

\bibitem{Hunt2016-vl}
Hunt TA, \emph{Mol Simul}, \textbf{42}, 347 (2016).

\bibitem{O-Connor2018-mh}
O'Connor TC, Alvarez NJ, Robbins MO, \emph{Phys Rev Lett}, \textbf{121}, 047801
  (2018).

\bibitem{OConnor2019-po}
O'Connor TC, Hopkins A, Robbins MO, \emph{Macromolecules}, \textbf{52}, 8540
  (2019).

\bibitem{O-Connor2020-ez}
O'Connor TC, Ge T, Rubinstein M, Grest GS, \emph{Phys Rev Lett}, \textbf{124},
  027801 (2020).

\bibitem{Murashima2021-wh}
Murashima T, Hagita K, Kawakatsu T, \emph{Macromolecules}, \textbf{54}, 7210
  (2021).

\bibitem{Murashima2022-kd}
Murashima T, Hagita K, Kawakatsu T, \emph{Macromolecules}, \textbf{55}, 9358
  (2022).

\bibitem{Jana2025-sh}
Jana S, Dalal IS, \emph{Macromol Theory Simul} (2025).

\bibitem{Jiang2007-rf}
Jiang W, Huang J, Wang Y, Laradji M, \emph{J Chem Phys}, \textbf{126}, 044901
  (2007).

\bibitem{Koide2023-ao}
Koide Y, Goto S, \emph{Soft Matter}, \textbf{19}, 4323 (2023).

\bibitem{Koide2022-bp}
Koide Y, Goto S, \emph{J Chem Phys}, \textbf{157}, 084903 (2022).

\bibitem{Pan2002-oq}
Pan G, Manke CW, \emph{J Rheol}, \textbf{46}, 1221 (2002).

\bibitem{evans_morriss_2008}
Evans DJ, Morriss GP, \emph{``Statistical Mechanics of Nonequilibrium
  Liquids''}, (2008), Cambridge University Press, Cambridge.

\bibitem{Daivis2006-pq}
Daivis PJ, Todd BD, \emph{J Chem Phys}, \textbf{124}, 194103 (2006).

\bibitem{Todd2007-ps}
Todd BD, Daivis PJ, \emph{Mol Simul}, \textbf{33}, 189 (2007).

\bibitem{Semaev2001-sn}
Semaev I, in \emph{``Cryptography and Lattices''}, Silverman JH (Ed), 181
  (2001), Springer, Berlin, Heidelberg.

\bibitem{Todd2000-rh}
Todd BD, Daivis PJ, \emph{J Chem Phys}, \textbf{112}, 40 (2000).

\bibitem{Nicholson2016-aj}
Nicholson DA, Rutledge GC, \emph{J Chem Phys}, \textbf{145}, 244903 (2016).

\bibitem{Dobson2016-zk}
Dobson M, Fox I, Saracino A, \emph{J Comput Phys}, \textbf{315}, 211 (2016).

\bibitem{Groot1997-je}
Groot RD, Warren PB, \emph{J Chem Phys}, \textbf{107}, 4423 (1997).

\bibitem{Murashima2018-te}
Murashima T, Hagita K, Kawakatsu T, \emph{Nihon Reoroji Gakkaishi},
  \textbf{46}, 207 (2018).

\bibitem{Irving1950-lc}
Irving JH, Kirkwood JG, \emph{J Chem Phys}, \textbf{18}, 817 (1950).

\bibitem{Liu2015-yj}
Liu MB, Liu GR, Zhou LW, Chang JZ, \emph{Arch Comput Methods Eng}, \textbf{22},
  529 (2015).

\bibitem{Gupta2000-ad}
Gupta RK, Nguyen DA, Sridhar T, \emph{Phys Fluids}, \textbf{12}, 1296 (2000).

\bibitem{uneyama2025radius}
Uneyama T, \emph{Nihon Reoroji Gakkaishi}, \textbf{53}, 11 (2025).

\bibitem{Zhang2015-oi}
Zhang Y, Otani A, Maginn EJ, \emph{J Chem Theory Comput}, \textbf{11}, 3537
  (2015).

\bibitem{Panoukidou2021-gz}
Panoukidou M, Wand CR, Carbone P, \emph{Soft Matter}, \textbf{17}, 8343 (2021).

\bibitem{Neelov2002-cp}
Neelov IM, Adolf DB, Lyulin AV, Davies GR, \emph{J Chem Phys}, \textbf{117},
  4030 (2002).

\bibitem{Stoltz2006-ht}
Stoltz C, de~Pablo JJ, Graham MD, \emph{J Rheol}, \textbf{50}, 137 (2006).

\bibitem{Prabhakar2017-no}
Prabhakar R, Sasmal C, Nguyen DA, Sridhar T, Prakash JR, \emph{Phys Rev
  Fluids}, \textbf{2}, 011301 (2017).

\bibitem{Ng1996-ed}
Ng SL, Mun RP, Boger DV, James DF, \emph{J Non-Newtonian Fluid Mech},
  \textbf{65}, 291 (1996).

\bibitem{Kim2019-xr}
Kim SG, Lee HS, \emph{Macromolecules}, \textbf{52}, 9585 (2019).

\bibitem{rubinstein2003polymer}
Rubinstein M, Colby RH, \emph{``Polymer Physics''}, (2003), Oxford University
  Press, Oxford.

\bibitem{Koide2025-zr}
Koide Y, Ishida T, Uneyama T, Masubuchi Y, \emph{Macromolecules}, \textbf{58},
  12039 (2025).

\bibitem{Hinch1977-lj}
Hinch EJ, \emph{Phys Fluids}, \textbf{20}, S22 (1977).

\bibitem{Fuller1981-mm}
Fuller GG, Leal LG, \emph{J Non-Newtonian Fluid Mech}, \textbf{8}, 271 (1981).

\bibitem{Magda1988-ha}
Magda JJ, Larson RG, Mackay ME, \emph{J Chem Phys}, \textbf{89}, 2504 (1988).

\bibitem{Prabhakar2016-ba}
Prabhakar R, Gadkari S, Gopesh T, Shaw MJ, \emph{J Rheol}, \textbf{60}, 345
  (2016).

\end{thebibliography}
\end{document}